# Electron-hole collision limited transport in charge-neutral bilayer graphene

Youngwoo Nam, Dong-Keun Ki, David Soler-Delgado, and Alberto F. Morpurgo*

Department of Quantum Matter Physics (DQMP) and Group of Applied Physics (GAP), University of Geneva, 24 Quai Ernest-Ansermet, CH1211 Genéve 4, Switzerland

*e-mail: Alberto.Morpurgo@unige.ch

**Ballistic transport occurs whenever electrons propagate without collisions deflecting their trajectory. It is normally observed in conductors with a negligible concentration of impurities, at low temperature, to avoid electron-phonon scattering. Here, we use suspended bilayer graphene devices to reveal a new regime, in which ballistic transport is not limited by scattering with phonons or impurities, but by electron-hole collisions. The phenomenon manifests itself in a negative four-terminal resistance that becomes visible when the density of holes (electrons) is suppressed by gate-shifting the Fermi level in the conduction (valence) band, above the thermal energy. For smaller densities transport is diffusive, and the measured conductivity is reproduced quantitatively, with no fitting parameters, by including electron-hole scattering as the only process causing velocity relaxation. Experiments on a trilayer device show that the phenomenon is robust and that transport at charge neutrality is governed by the same physics. Our results provide a textbook illustration of a transport regime that had not been observed previously and clarify the nature of conduction through charge-neutral graphene under conditions in which carrier density inhomogeneity is immaterial. They also demonstrate that transport can be limited by a fully electronic mechanism, originating from the same microscopic processes that govern the physics of Dirac-like plasmas.**

Ever since Sharvin's pioneering work[1], the occurrence of ballistic transport in metallic conductors has been exploited to investigate the electronic properties of solids. In the quasi-classical regime, for instance, magnetic focusing experiments have allowed probing electron-dynamics and the shape of the Fermi surface in ultra-pure crystals of different metals[2], in semiconducting heterostructures[3,4], and –more recently– in graphene-based systems[5,6]. In the quantum regime, when the electron wavelength and the conductor size are comparable, ballistic motion normally leads to conductance quantization and allows probing transport through individual quantum channels[7,8]. In all cases, the observation of ballistic transport requires the elastic mean free path determined by collisions of electrons with impurities to be longer than the system size, and the rate of inelastic processes such as phonon scattering to be sufficiently small. Electron-electron collisions are less detrimental, as they only slowly de-collimate a focused beam of ballistic electrons influencing their trajectories gradually, with effects that usually become relevant at rather high temperature[6,9].

Here we show that when both electrons and holes are simultaneously present the situation is drastically different, and that electron-hole collisions play a dominant role. Our work relies on multi-terminal suspended bilayer graphene device (Fig. 1a) of the highest possible quality[10], with carrier density inhomogeneity smaller than $10^9$ cm$^{-2}$, as inferred from the gate-dependent transport properties (Fig. 1b, c). Earlier, we had reported that in these devices ballistic transport manifests itself in a negative four-terminal resistance[10], but had not analyzed thoroughly the conditions for its occurrence. To start addressing this point, we look at the evolution of the onset of the ballistic regime –marked by the crossing point from positive to negative four-terminal resistance– with carrier density ($n$) and temperature ($T$), shown in Fig. 1d.

Fig. 1e shows that the crossing point varies rapidly upon varying $T$, so that for larger $T$ larger carrier densities are needed for electron motion to become ballistic. This trend is indicative of a microscopic scattering mechanism that causes the electronic mean free path to depend strongly on $n$ and $T$. Since in the experiments $n$ is varied between a few $10^9$ cm$^{-2}$ and $2 \times 10^{11}$ cm$^{-2}$ –well above the range associated to carrier density inhomogeneity– and $T$ between 12 and 100 K –below the temperature where electron-phonons scattering sets in[11,12]– neither density inhomogeneity nor phonon scattering can account for large changes in mean free path. Furthermore, because the density of states is constant in bilayers, a pronounced dependence of the screening length on $n$ and $T$ can also not be invoked. To gain insight, we compare the Fermi energy $E_F$ to the thermal energy $k_B T$ at the onset of negative multi-terminal resistance ($E_F$ is obtained from the gate-induced electron density, using the known effective mass[13-16], $m^* = 0.033\ m_0$, with $m_0$ the free electron mass). The data in Fig. 1d show if $n$ is increased further, the negative four-terminal resistance first increases (in modulus) and eventually saturates when $E_F$ exceeds 2-3 times $k_B T$. This behavior is summarized in Fig. 1e, which shows that the multi-terminal resistance vanishes when $E_F$ is comparable to $k_B T$, so that transport is diffusive if $E_F \ll k_B T$ and becomes ballistic if $E_F \gg k_B T$. (the precise behavior in the crossover regime is determined by the device geometry and dimensions, since the occurrence of ballistic transport requires the mean-free path to be longer than the device size).

Finding that scattering is drastically suppressed if $E_F$ significantly exceeds $k_B T$ suggests that electron-hole collisions play a relevant role because, in a zero-gap semiconductor like bilayer graphene, the density of minority carriers (holes, if the Fermi energy is in the conduction bands; electrons, in the opposite case) is suppressed exponentially as $e^{-\frac{E_F}{K_B T}}$. Therefore, if collisions on minority carriers were the relevant scattering process, a strong $n$ and $T$ dependence of the mean free path should be expected. As it has long been known that –in contrast to electron-electron or hole-hole collisions– electron-hole scattering can effectively cause velocity relaxation[17-21] (see Fig. 2a-c), this hypothesis represents a realistic scenario. Indeed, it has been proposed theoretically in monolayer graphene that electron-hole scattering determine the transport properties near charge neutrality[22-26] (i.e., in the region of interest here, where $k_B T > E_F$). To validate a scenario based on electron-hole scattering we analyze quantitatively transport in the diffusive regime at low $n$, to check whether the density and temperature dependence of the electron-hole collision rate can reproduce quantitatively the experimentally measured conductivity.

To analyze the data, we follow a phenomenological approach that takes into account the specific aspects of our system: the band structure of bilayer graphene, the inelastic scattering due to carrier-carrier interaction at charge neutrality, as well as the fact that the experiments are carried out on a small size device on the verge of ballistic transport. The goal is to reproduce the behavior observed in the charge neutrality region ($k_B T \geq E_F$), in which electrons and holes coexist, the multi-terminal resistance is positive and transport is still in the diffusive regime. To this end, we need an expression for the experimentally measurable quantities (the conductivity $\sigma$ as a function of $T$ and $n = C_g(V_g - V_{CNP})$, where $C_g$ is the capacitance to the gate electrode) that can be compared to the data.

To make the discussion concrete, we describe transport within a quasi-classical approach, which is possible because the characteristic wavelength of electrons and holes –the thermal wavelength near charge neutrality, since $E_F < k_B T$– is always sufficiently smaller than the device dimensions (the thermal wavelength is ~200 nm at 10 K, and shorter at larger $T$; as we discuss below, data and theory will be compared for temperatures ranging approximately between 12 and 100 K). Within this approach, the conductivity is given by the sum of the electron and hole contributions, expressed in terms of the respective density $n_e$ and $n_h$ and scattering rates $\tau_e^{-1}$ and $\tau_h^{-1}$. That is: $\sigma = n_e e \mu_e + n_h e \mu_h$ where $\tau_e$ and $\tau_h$ represent the scattering time responsible for relaxation of the velocity of charge carriers, i.e., the times that enter the expression for the electron and hole mobility, $\mu_{e/h} = \frac{e \tau_{e/h}}{m^*}$. Using the known constant density of states of bilayer graphene $g = \frac{2m^*}{\pi \hbar^2}$, it is straightforward to calculate the dependence of $n_e$ and $n_h$ on $T$ and $E_F$ (e.g., for the case in which $E_F$ is in the conduction band, $n_e = g E_F + g k_B T ln\left(1 + e^{-\frac{E_F}{k_B T}}\right)$ and $n_h = g k_B T ln\left(1 + e^{-\frac{E_F}{k_B T}}\right)$) and to show that $E_F = \frac{\pi \hbar^2}{2 m^*} n$ is independent of temperature. What remains to be done is to determine the scattering rates $\tau_e^{-1}$ and $\tau_h^{-1}$ as a function of $n$ and $T$.

The rate of electron scattering processes causing velocity relaxation can be written as $\frac{1}{\tau_e} = \frac{1}{\tau} + \Gamma_{e-h}$, where $\tau^{-1}$ is the rate for scattering on impurities and phonons, and $\Gamma_{e-h}$ the rate for scattering of electrons on holes. Analogously, for holes we write $\frac{1}{\tau_h} = \frac{1}{\tau} + \Gamma_{h-e}$. When as $k_B T > E_F$ (i.e., at charge neutrality), the total scattering rate $\Gamma$ for collisions between charge carriers (irrespective of whether they are electrons or holes) in monolayer graphene is given by $\Gamma = C \frac{k_B T}{\hbar}$, with $C \sim 1$[22-26]. This relation is also expected to hold more generally (see for instance Ref. [27]). As such, we assume that it applies also to our bilayer graphene samples. Since velocity relaxation is caused by collisions between electrons and holes, and not by electron-electron or hole-hole scattering, the scattering rates $\Gamma_{e-h}$ and $\Gamma_{h-e}$ are related to $\Gamma$ as $\Gamma_{e-h} = \Gamma \frac{n_h}{n_e+n_h}$ and $\Gamma_{h-e} = \Gamma \frac{n_e}{n_e+n_h}$. Using these expressions all quantities needed to calculate $\sigma$ can be expressed in terms of $n$ and $T$, except for the scattering time $\tau$. Having in mind that electron-hole scattering dominates we assume that the time $\tau$ is longer than the time taken by charge carriers to traverse the device and can be neglected near charge neutrality, which is also why ballistic transport can be observed in the experiments when the density of minority carriers is sufficiently small (i.e., we assume that momentum relaxation only occurs in the contacts and at the edges of the device). Under this assumption the expression for the conductivity reads:

$$\sigma(n,T) = n_e e \mu_e + n_h e \mu_h = \frac{1}{C} \frac{\hbar}{k_B T} \frac{e^2}{m^*} (n_e + n_h) \frac{n_e^2 + n_h^2}{n_e n_h}, \qquad (1)$$

Eq. (1) is expected to hold as long as $k_B T > E_F$ since otherwise the density of minority carriers becomes too small to cause sizable scattering rates, motion becomes ballistic, and transport cannot be described in terms of conductivity (note that $T$ should also remain sufficiently low not to have effects due to phonon scattering, which in bilayer graphene means $T \leq 100$ K)[11,12].

To compare Eq. (1) with the data, we look first at the normalized conductivity

$$\frac{\sigma(n,T)}{\sigma(0,T)} = \frac{\pi \hbar^2}{8 k_B T m^* \ln(2)} \frac{(n_e+n_h)(n_e^2+n_h^2)}{n_e n_h}, \qquad (2)$$

which does not contain any free parameter (both the constant $C$ and the device dimensions –i.e., the precise values of width and length used to calculate the conductivity from the experimental data– cancel out when looking at $\frac{\sigma(n,T)}{\sigma(0,T)}$). The conductivity is obtained from multi-terminal measurements performed by sending current from contact 1 to 2 and measuring voltage between contact 4 and 3 (see Fig. 1a), i.e. probing transport along the "long" device direction, in which the effects of ballistic transport are less pronounced (raw data of resistance versus gate voltage are shown in the supplementary information). Fig. 2d shows that the agreement with the data is excellent throughout the range investigated. For the device shown here this corresponds to 12 K $\leq T \leq$ 40 K and $10^9 \leq n \leq 10^{11}$ cm$^{-2}$; in other devices shown in the supplementary information, measurements have been done –and found to agree with theory– up to ~ 100 K. The analysis is confined to $T \geq 12$ K, because at lower temperature an interaction-induced broken symmetry state occurs in suspended bilayer graphene near charge neutrality[28-30], a gap opens near charge neutrality, and the device become insulating (*i.e.*, for $T < 12$ K transport is dominated by the opening of the gap and not by e-h scattering; data for $T < 12$ K illustrating the behavior of transport in the broken symmetry insulating state is included in the supplementary information). The double-logarithmic plot of the data again indicates an impressively good agreement as $n$ is varied over two orders of magnitude as shown in Fig. 2e. The only flexibility in this comparison is given by the precise value of the effective mass. In the literature[13-16] $m^*$ is reported to range from 0.025 to 0.04 $m_0$, and we use the most commonly cited value $m^* = 0.033$ $m_0$. Data measured on four different devices –two four-terminal devices and two two-terminal devices– were found to all exhibit an excellent agreement, with nearly identical values of effective mass ($m^* = 0.031$-$0.034$ $m_0$, data from other devices are shown in the supplementary information).

Interestingly, Eq. (2) indicates that $\frac{\sigma(n,T)}{\sigma(0,T)}$ depends on $n$ and $T$ only through the combination $E_F/k_BT$. It should then be expected that when plotting the data versus $E_F/k_BT$ all curve collapse on top of each other irrespective of the temperature and density range in which the measurements are done, as long as Eq. (2) holds. Fig. 2f shows that this is indeed the case. Not only all the curves exhibit a very good collapse and the agreement with theory excellent, but also deviations become increasingly more pronounced as $E_F/k_BT$ is increased, as expected. In this device, the agreement is satisfactory for $E_F/k_BT$ approaching 1; in other devices (see for instance Fig. 3f) deviations start for somewhat smaller values of $E_F/k_BT$, albeit still of the order of 1. These small differences originate from the fact that the details of the crossover from diffusive to ballistic transport are device dependent, as they are sensitive –for instance– to the precise device geometry and dimensions. Finding such a good collapse of the measurements when the data are plotted versus $E_F/k_BT$ provides additional evidence for the validity of Eq. (2) and, accordingly, for the dominant role of electron-hole scattering as the process that limits transport in the vicinity of charge neutrality.

Graphene bilayers are ideal systems to carry out these investigations, because their relatively large density of states (as compared to monolayers) makes them drastically less sensitive to carrier density inhomogeneity. Specifically, in monolayers an inhomogeneity of ~ $10^9$ cm$^{-2}$ leads to fluctuations in Fermi energy in excess of 50 K, so that measurement could only be performed at higher temperatures. However, for $T \sim 100$ K phonons start also to influence transport[31], making the temperature range available in monolayers quite narrow[32]. Conversely, all considerations made for bilayers should hold true for trilayers, whose band structure consists of a parabolic bilayer-like band and of a linear Dirac band (Fig. 3c)[33-37]. Since the density of states of the linear band is negligible at the energies relevant for our work (the carrier density contributed by the linear band is less than 1%), it is reasonable to expect that transport is dominated by the quadratic, bilayer-like band. If so, Eq. (1) and (2) above hold, as long as the appropriate value of effective mass is used that, according to theory, is $\sqrt{2}$ times larger than in bilayers[37] (giving in our case $m^*_{tri} = 0.047$ $m_0$). Fig. 3d shows the experimental data for the normalized conductivity $\frac{\sigma(n,T)}{\sigma(0,T)}$, together with the result of Eq. (2) using $m^* = 0.06$ $m_0$. The agreement is once again excellent and it persists as $n$ is varied over two orders of magnitude (see double-logarithmic plot in Fig. 3e). The trilayer effective mass extracted from the experiments is larger than that of bilayers, as anticipated by theory, and deviates by only 20% from the theoretical value. Just as for bilayers, also for trilayers all curves collapse on top of each other when plotted versus $E_F/k_BT$ (Fig. 3f). Observing such a good agreement between data and theory in a trilayer device is particularly important, as it represents an independent confirmation of the validity of Eq. (2) and, accordingly, of the underlying phenomenological description of transport near charge neutrality (i.e., the experiments on the trilayer device provide an independent confirmation of the dominant role of electron-hole scattering near charge neutrality).

We now verify whether Eq. (1) also correctly reproduces the absolute value of the measured quantities –conductivity and mobility– and not only the normalized ones. In making this comparison, it should be borne in mind that although $C \sim 1$, the precise value is not exactly known and $C$ is expected to be weakly dependent on $T$ and $n^{25}$. Additionally, there is an uncertainty associated with the device geometry, due to the small dimensions that can result in a not perfectly uniform current density distribution. Therefore, it should be considered as satisfactory if –at a quantitative level– Eq. (3) agrees with the data within a factor of 2-3. We first look at the temperature dependence of the mobility at charge neutrality. In the past, analysis performed on high-quality suspended graphene devices have repeatedly assumed the mobility at charge neutrality to be temperature independent[38,39], illustrating how the relevance of e-h scattering was not properly appreciated in the interpretation of earlier experiments. Indeed, our description of transport leading to Eq. (1) predicts the mobility at charge neutrality to be proportional to *1/T*,

$$\mu(n=0,T) = \frac{\sigma\ (n=0,T)}{e[n_e(n=0,T)+n_h(n=0,T)]} = \frac{2}{C}\frac{\hbar}{k_BT}\frac{e}{m^*}, \tag{3}$$

(at $n = 0$, electrons and holes have identical density $n_e = n_h = \frac{2m^*}{\pi\hbar^2}k_BT ln(2)$), which is a direct manifestation of the electron-hole scattering mechanism that is responsible for limiting transport as long as $E_F/k_BT \ll 1$. The comparison with experimental data is shown in Fig. 4a for bilayer graphene, for the entire temperature range over which each one of the four different devices studied have been measured. Except for one device, the predicted mobility value does indeed agree with the measurements within a factor of 2 or better. For the trilayer graphene device as well the measured absolute value of the mobility at charge neutrality agrees within better than a factor of 2 with the theoretically expected one, throughout the entire temperature range investigated (see Fig. 4b).

As for the conductivity, a direct consequence of the dominant role of electron-hole scattering is that at charge neutrality:

$$\sigma(n=0,T) = \frac{16 ln(2)}{C}\left(\frac{e^2}{h}\right), \tag{4}$$

It follows that only a very weak temperature dependence of $\sigma(n = 0, T)$ should be expected, originating from the temperature dependence of the coefficient C, which may include –for example – logarithmic corrections to the *1/T* scattering rate. Fig. 4c and d indeed show that –except for the same bilayer device for which the mobility exhibits a larger deviation from theory– for all bilayer devices and for the trilayer device the conductivity at charge neutrality is very weakly dependent on temperature (the variations are certainly within the range that can be explained by the expected logarithmic corrections to the coefficient C) and agrees within better than a factor of 2 with the theoretically predicted value. Since, once again, no parameter can be varied to optimize the theoretical predictions, the agreement we find between Eq. (1) and the absolute value of the measured carrier mobility and conductivity further strengthens our conclusions about the role of electron-hole scattering.

We conclude that electron-hole scattering determines transport near charge neutrality in sufficiently clean zero-gap semiconductors, such as suspended bilayer and trilayer graphene. Whereas the underlying theoretical concepts are long known, the interpretation of earlier experimental studies addressing the role of electron-hole scattering processes have often not been uncontroversial[20]. In the suspended graphene devices discussed here the situation is different because all microscopic system parameters are known in great detail, transport can be gate-tuned, and the occurrence of ballistic transport at sufficiently high carrier density shows that electron-hole scattering is the only process causing velocity relaxation on the length scale of our devices. Besides illustrating experimentally an interesting new mechanism limiting ballistic transport, these findings have implications for the physics of graphene. They identify the processes determining the intrinsic transport properties in graphene near charge neutrality in the absence of disorder, for which different mechanisms have been proposed theoretically[13,40-44]. In particular our results establish that, at any finite temperature, conduction is due to thermal activation from the valence to conduction band, which fixes the density of charge carriers, and is limited by electron-hole collisions, which determine the carrier mobility.

Even more importantly, these results illustrate how electrical transport can be limited by an exclusively electronic mechanism, which is why an expression for the conductivity exhibiting virtually perfect quantitative agreement with the data can be derived. Such an agreement validates the expression for the scattering rate between charge carriers, $\Gamma = C\frac{k_B T}{\hbar}$, which is believed to hold for many metallic conductors whose behavior is not captured by the Fermi liquid paradigm, and that should be described in terms of quantum criticality[27]. This is of particular interest, because – at charge neutrality– these same inter-particle collisions are the processes bringing Dirac-like plasmas into internal equilibrium, and creating the conditions needed to reach the hydrodynamic transport regime. So far, the hydrodynamic regime has been observed experimentally in high-quality graphene on hBN at high carrier density, away from charge neutrality, under conditions in which only one type of charge carrier is present[45,46], but not near charge neutrality –when charge inhomogeneity poses more stringent constraints. It is nevertheless in the charge neutrality region that a violation of the Wiedemann-Franz law[32] has been reported, for which a quantum critical scenario based on a hydrodynamic description has been proposed[44]. It will likely require more investigations to determine whether or not quantum criticality is the correct framework to interpret the violation of the Wiedemann-Franz law and possibly other experiments in charge-neutral graphene. Nevertheless, our results do show that, in the charge neutrality region, the dynamics of electrons and holes is governed by an inelastic scattering rate $\Gamma \sim \frac{k_B T}{\hbar}$, which conforms to the expected behavior of a quantum critical system. They also show that in this regime the conditions for needed for a description in hydrodynamic terms of a two-component electron-hole plasma are met in sufficiently clean devices.

## Methods

Fig. 1a and 3a show optical images of suspended graphene devices. The devices are made using polydimethylglutarimide (PMGI)-based lift-off resist (LOR 10A, MicroChem) as a sacrificial layer[47]. LOR resist (1 μm-thick) is spun on heavily doped Si substrate serving as a back-gate, covered with 285 nm-thick $SiO_2$ and graphene flakes from natural graphite are mechanically exfoliated onto it. The thickness of the graphene layers is initially identified by means of the observed optical contrast and subsequently confirmed with transport measurement (quantum Hall measurements, as well as temperature and gate voltage dependent conductivity measurements; for the trilayer device these measurements indicate also that the stacking is of the Bernal type). Electrodes (10 nm Ti/ 70 nm Au) are patterned using a conventional electron-beam lithography followed by electron-beam metal evaporation. The shape of the suspended channel is defined by etching graphene (white area enclosed by dashed lines in the optical images) with an oxygen plasma. Finally, graphene suspension is realized by selectively removing LOR resist underneath. The devices have been characterized in great detail by investigating transport in the quantum Hall regime[10,48-50], and the observed behavior conforms to the one expected and found no evidence of effects due to strain (which could be expected due to the force acting between the gate and the suspended layer).

Before starting transport measurements, the devices are biased up to large voltage (~ 2 V) at 4 K in vacuum, to perform a so-called current annealing process, necessary to remove adsorbates adhering to graphene. Such a current annealing is performed by repeatedly sweeping the applied bias until when the peak at charge neutrality in the resistivity-vs-gate voltage curve becomes extremely narrow, corresponding to a charge inhomogeneity level of $\delta n \sim 10^9$ $cm^{-2}$, or smaller. Such a high device quality is necessary to prevent charge inhomogeneity from masking the transport regime dominated by electron-hole scattering. The yield of successfully current annealed devices of this quality is low, which makes the experiments time consuming, but it nevertheless allows the realization of a number of devices sufficient to check the reproducibility of the experimental results (a total of five different devices have been studied, exhibiting excellent quantitative agreement with the expression for the conductivity discussed above).

All transport measurements are performed using a conventional low-frequency lock-in technique. In multi-terminal devices two types of measurement configurations are employed, with current and voltage probes placed either along the long-channel direction (with reference to the contact scheme shown in Fig. 1a, *I*: 1-2, *V*: 4-3) or along the short-channel direction (*I*: 1-4, *V*: 2-3)[10]. The latter configuration enables the observation of negative resistance demonstrating the occurrence of ballistic transport. In the manuscript, except the measurement showing the negative resistance (Fig. 1d), all measurements performed to extract the density and temperature dependent conductivity were carried out in the configuration probing transport along the long channel direction, in which the multi-terminal resistance is positive.

## Acknowledgements

We gratefully acknowledge A. Ferreira for continued technical support of the experiments. We are also grateful to D. Abanin, V. Fal'ko, T. Giamarchi, L.S. Levitov, M. Müller, M. Polini, J. Song, D. Valentinis, D. van der Marel, and J. Wallbank for very helpful discussions. Financial support from the Swiss National Science Foundation, the NCCR QSIT, and the EU Graphene Flagship Project are also gratefully acknowledged.

**Authors Contributions**

Y.N, D.K.K, and D.S.D fabricated devices and Y.N and D.K.K performed measurements. A.F.M derived the expression for the conductivity and supervised the analysis of the data done by Y.N. Y.N, D.K.K, and A.F.M wrote the manuscript. All authors discussed the results and contributed to their interpretation.

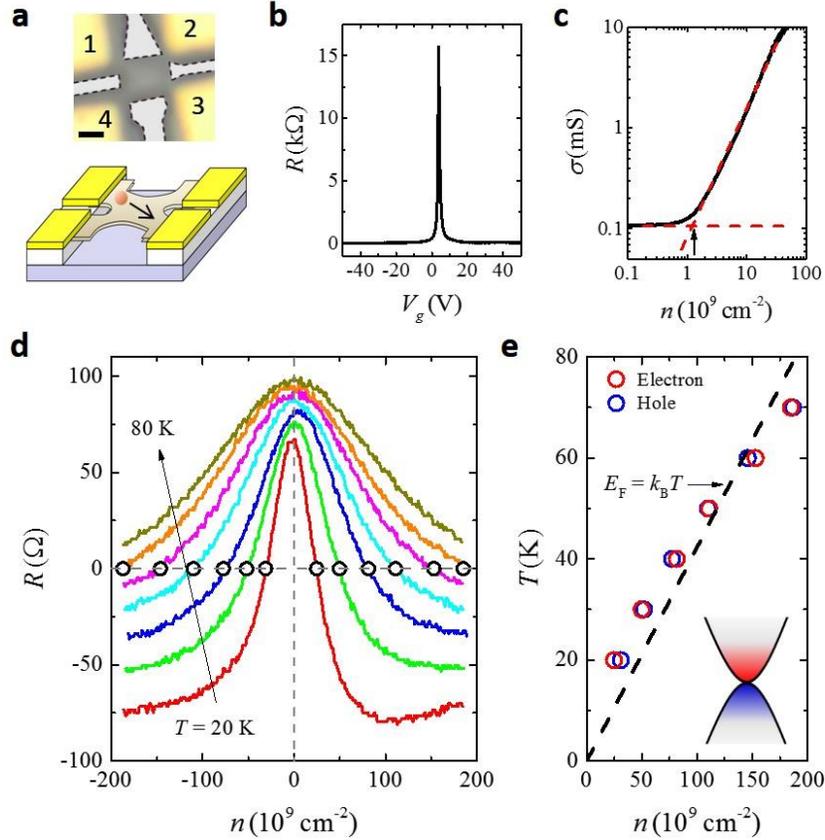

**Figure 1 – Ballistic transport in multi-terminal suspended bilayer devices. a.** Optical microscope image (top) and schematics of suspended device structure (bottom). The electrical contacts are labelled by the corresponding numbers. The scale bar is 1 μm. **b.** Gate-voltage dependence of the resistance measured at $T = 4$ K, showing a very sharp peak around charge neutrality. **c.** Double-logarithmic plot of the conductance as a function of carrier density extracted from **b**. The carrier density $n$ is linearly proportional to the gate voltage $V_g$ and the coefficient $\alpha = n/V_g \sim 5.3 \times 10^9$ cm$^{-2}$/V is obtained from the analysis of quantum Hall measurements[10]. A very low charge inhomogeneity $\delta n \sim 1$–$2 \times 10^9$ cm$^{-2}$ is estimated by looking at the crossover from a constant to a $V_g$-dependent conductance. **d.** Four-terminal resistance versus carrier density at various temperatures from 20 to 80 K with a 10 K step (current is sent from contact 1 to 4; voltage is measured between contacts 2 and 3). Upon increasing carrier density, the multi-terminal resistance becomes negative because of collimation of the carriers injected from contact 1, indicating the occurrence of ballistic transport. The onset of negative resistance, marked by empty circles, strongly depends on $n$ and $T$, as summarized in panel **e**. The dashed line in this panel corresponds to $E_F = k_B T$, indicating that ballistic transport dominates when the Fermi energy is larger than the thermal energy (the Fermi energy is obtained by using the effective mass of bilayer graphene, $m^* = 0.033\, m_0$, to calculate the density of states –see main text). The lower inset represents schematically the band structure of bilayer graphene that –at charge neutrality at finite temperature– is populated with thermally excited electrons and holes.

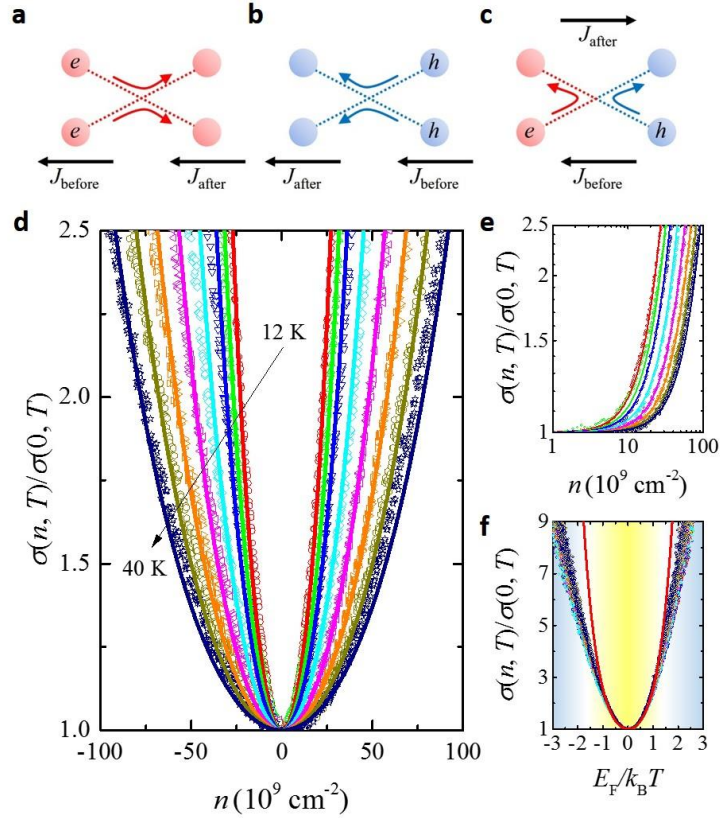

**Figure 2 – Electron-hole scattering in bilayer graphene. a-c.** Schematic representation of electron-electron, hole-hole, and electron-hole scattering processes. As a result of momentum conservation, electron-electron and hole-hole collisions conserve current. That is not the case for electron-hole collisions, which cause current (and, therefore, velocity) relaxation and limit transport. **d.** The empty symbols represent the normalized conductivity measured by sending current from contact 1 to 2 and probing voltage drop between contacts 4 and 3 (see scheme in Fig. 1a) as a function of $n$ at various temperatures, $T$ = 12, 14, 16, 20, 25, 30, 35, 40 K. The continuous lines of the corresponding color are obtained by plotting Eq. (2) with $m^* = 0.033\ m_0$. They show an excellent agreement with the experiments upon varying $n$ over two orders of magnitude, as illustrated by the double-logarithmic plot in **e**. **f.** When plotting the normalized conductivity as a function of $E_F/k_BT$, all data for different temperatures collapse onto a single curve, which –as long as $E_F/k_BT < 1$ (yellow shaded region)– is in excellent agreement with the prediction of Eq. (2), represented by the red continuous curve. For $E_F/k_BT \gg 1$ (light blue shaded region) deviations occurs because transport enters the ballistic regime (see Fig. 1d) and cannot be described in terms of conductivity.

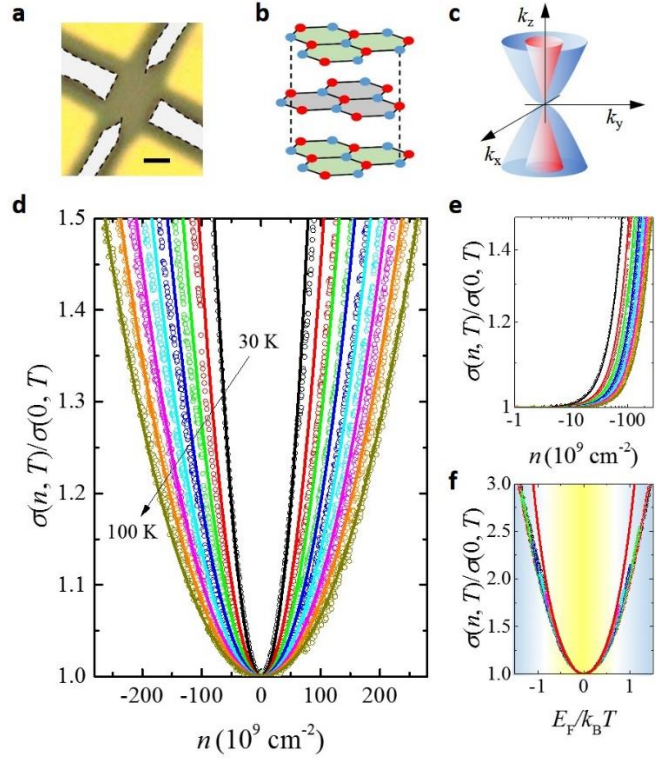

**Figure 3 – Electron-hole scattering in Bernal stacked trilayer graphene. a.** Optical microscope image of the multi-terminal suspended trilayer device investigated here. The scale bar is 1 μm. **b.** Schematic representation of the structure of Bernal-stacked trilayer graphene and of its bands near charge neutrality (**c**). **d.** The empty symbols represent the normalized conductivity measured as a function of $n$, as the temperature $T$ is varied between 30 and 100 K in 10 K steps. We look at measurements for $T \geq 30$ K only, because for smaller temperature values electron-electron interactions have an important effect for Bernal-stacked trilayers (i.e., not only for ABC trilayers) and modify the density of states by opening a gap in the bilayer band (see e.g., Refs. [49,50]; direct experimental evidence will be discussed elsewhere). The continuous lines of the corresponding color are obtained by plotting Eq. (2) with $m^* = 0.06\, m_0$, and show an excellent agreement with the experiments. As for bilayers, also for trilayers, the agreement is excellent as $n$ is varied over two orders of magnitude, as shown by the double-logarithmic plot of the data in **e**. The gate capacitance needed to relate $n$ and $V_g$ ($\alpha = n/V_g \sim 5.5\times10^9$ cm$^{-2}$/V) is determined from measurements of the quantum Hall effect, as detailed in the supplementary information. **f.** The normalized conductivity for all different temperature collapse onto a single curve when plotted as a function of $E_F/k_BT$, in virtually perfect agreement with Eq. (2) (red line) for $E_F/k_BT < 0.5$.

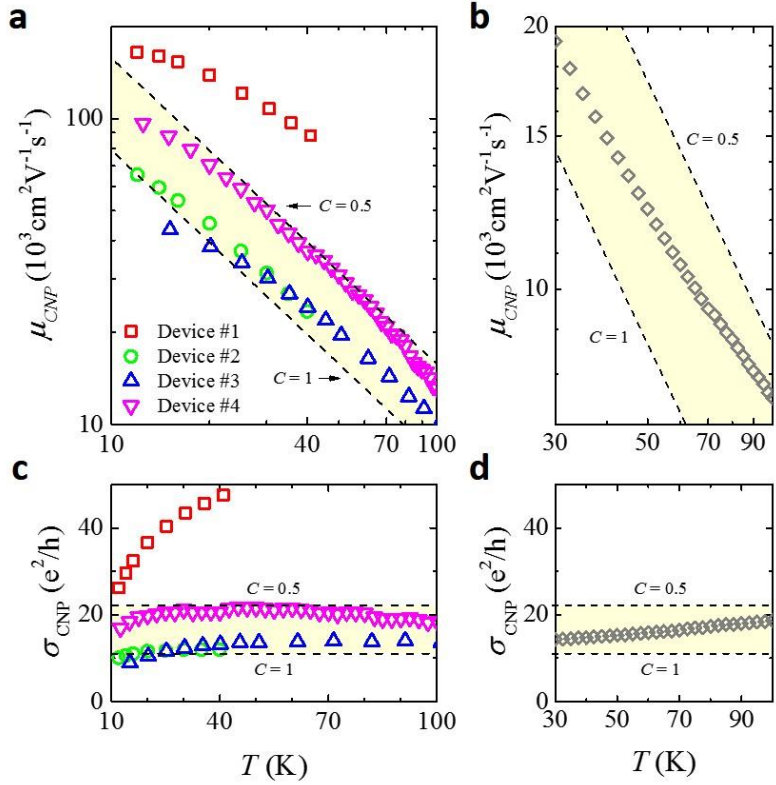

**Figure 4 – Temperature dependence of the mobility and conductivity at charge neutrality. a, c.** Double-logarithmic plot of mobility versus temperature (a) and linear plot of conductivity versus temperature (c) for four different bilayer graphene devices (device #1 and #2 are multi-terminal devices and device #3 and #4 are two-terminal devices). The dashed lines correspond to the mobility value expected from Eq. (3), with $C =1$ and $C = 0.5$. The data agree quantitatively with the theory within a factor of 2 except for one bilayer device. This indetermination is partly due to the unknown precise value of $C$ and partly to the small device geometry that causes a non-perfectly uniform current flow. **b, d.** Quantitative agreement with the same level of indetermination is observed for the trilayer device that we have investigated (the dashed lines correspond to Eq. (4) with $C = 1$ and $C = 0.5$, respectively). There are no adjustable parameters that can be varied to achieve this level of agreement.

# References

1. Sharvin, Y. V. A Possible Method for Studying Fermi Surfaces. *J. Exp. Theor. Phys.* **21**, 655 (1965).
2. Tsoi, V. S., Bass, J. & Wyder, P. Studying conduction-electron/interface interactions using transverse electron focusing. *Rev. Mod. Phys.* **71**, 1641-1693 (1999).
3. van Houten, H. *et al.* Coherent electron focusing with quantum point contacts in a two-dimensional electron gas. *Phys. Rev. B* **39**, 8556-8575 (1989).
4. Beenakker, C. W. J. & Vanhouten, H. Quantum Transport in Semiconductor Nanostructures. *Solid State Phys.* **44**, 1-228 (1991).
5. Taychatanapat, T., Watanabe, K., Taniguchi, T. & Jarillo-Herrero, P. Electrically tunable transverse magnetic focusing in graphene. *Nat. Phys.* **9**, 225-229 (2013).
6. Lee, M. *et al.* Ballistic miniband conduction in a graphene superlattice. *Science* **353**, 1526-1529 (2016).
7. van Wees, B. J. *et al.* Quantized conductance of point contacts in a two-dimensional electron gas. *Phys. Rev. Lett.* **60**, 848-850 (1988).
8. Wharam, D. A. *et al.* One-dimensional transport and the quantisation of the ballistic resistance. *J. Phys. C: Solid State Phys.* **21**, L209 (1988).
9. Predel, H. *et al.* Effects of electron-electron scattering on electron-beam propagation in a two-dimensional electron gas. *Phys. Rev. B* **62**, 2057-2064 (2000).
10. Ki, D.-K. & Morpurgo, A. F. High-quality multiterminal suspended graphene devices. *Nano Lett.* **13**, 5165-5170 (2013).
11. Ochoa, H., Castro, E. V., Katsnelson, M. I. & Guinea, F. Temperature-dependent resistivity in bilayer graphene due to flexural phonons. *Phys. Rev. B* **83**, 235416 (2011).
12. Laitinen, A. *et al.* Coupling between electrons and optical phonons in suspended bilayer graphene. *Phys. Rev. B* **91**, 121414 (2015).
13. Koshino, M. & Ando, T. Transport in bilayer graphene: Calculations within a self-consistent Born approximation. *Phys. Rev. B* **73**, 245403 (2006).
14. McCann, E. & Fal'ko, V. I. Landau-Level Degeneracy and Quantum Hall Effect in a Graphite Bilayer. *Phys. Rev. Lett.* **96**, 086805 (2006).
15. Gorbachev, R. V., Tikhonenko, F. V., Mayorov, A. S., Horsell, D. W. & Savchenko, A. K. Weak Localization in Bilayer Graphene. *Phys. Rev. Lett.* **98**, 176805 (2007).
16. Li, J. *et al.* Effective mass in bilayer graphene at low carrier densities: The role of potential disorder and electron-electron interaction. *Phys. Rev. B* **94**, 161406 (2016).
17. Baber, W. G. The contribution to the electrical resistance of metals from collisions between electrons. *Proc. R. Soc. London, A* **158**, 0383-0396 (1937).
18. Thompson, A. H. Electron-Electron Scattering in $TiS_2$. *Phys. Rev. Lett.* **35**, 1786-1789 (1975).
19. Kukkonen, C. A. & Maldague, P. F. Electron-Hole Scattering and the Electrical Resistivity of the Semimetal $TiS_2$. *Phys. Rev. Lett.* **37**, 782-785 (1976).
20. Gantmakher, V. F. & Levinson, I. B. Effect of Collisions between Current Carriers on Dissipative Conductivity. *Sov. Phys. JETP* **47**, 133-137 (1978).
21. Entin, M. V. *et al.* The effect of electron-hole scattering on transport properties of a 2D semimetal in the HgTe quantum well. *J. Exp. Theor. Phys.* **117**, 933-943 (2013).
22. González, J., Guinea, F. & Vozmediano, M. A. H. Marginal-Fermi-liquid behavior from two-dimensional Coulomb interaction. *Phys. Rev. B* **59**, R2474-R2477 (1999).
23. Sheehy, D. E. & Schmalian, J. Quantum Critical Scaling in Graphene. *Phys. Rev. Lett.* **99**, 226803 (2007).
24. Son, D. T. Quantum critical point in graphene approached in the limit of infinitely strong Coulomb interaction. *Phys. Rev. B* **75**, 235423 (2007).

25. Fritz, L., Schmalian, J., Müller, M. & Sachdev, S. Quantum critical transport in clean graphene. *Phys. Rev. B* **78**, 085416 (2008).
26. Kashuba, A. B. Conductivity of defectless graphene. *Phys. Rev. B* **78**, 085415 (2008).
27. Hartnoll, S. A. Theory of universal incoherent metallic transport. *Nat. Phys.* **11**, 54-61 (2015).
28. Weitz, R. T., Allen, M. T., Feldman, B. E., Martin, J. & Yacoby, A. Broken-Symmetry States in Doubly Gated Suspended Bilayer Graphene. *Science* **330**, 812-816 (2010).
29. Freitag, F., Trbovic, J., Weiss, M. & Schönenberger, C. Spontaneously Gapped Ground State in Suspended Bilayer Graphene. *Phys. Rev. Lett.* **108**, 076602 (2012).
30. Velasco Jr., J. *et al.* Transport spectroscopy of symmetry-broken insulating states in bilayer graphene. *Nat. Nanotech.* **7**, 156-160 (2012).
31. Castro, E. V. *et al.* Limits on Charge Carrier Mobility in Suspended Graphene due to Flexural Phonons. *Phys. Rev. Lett.* **105**, 266601 (2010).
32. Crossno, J. *et al.* Observation of the Dirac fluid and the breakdown of the Wiedemann-Franz law in graphene. *Science* **351**, 1058-1061 (2016).
33. Guinea, F., Castro Neto, A. & Peres, N. Electronic states and Landau levels in graphene stacks. *Phys. Rev. B* **73**, 245426 (2006).
34. Latil, S. & Henrard, L. Charge Carriers in Few-Layer Graphene Films. *Phys. Rev. Lett.* **97**, 036803 (2006).
35. Koshino, M. & Ando, T. Orbital diamagnetism in multilayer graphenes: Systematic study with the effective mass approximation. *Phys. Rev. B* **76**, 085425 (2007).
36. Partoens, B. & Peeters, F. Normal and Dirac fermions in graphene multilayers: Tight-binding description of the electronic structure. *Phys. Rev. B* **75**, 193402 (2007).
37. Koshino, M. Interlayer screening effect in graphene multilayers with ABA and ABC stacking. *Phys. Rev. B* **81**, 125304 (2010).
38. Mayorov, A. S. *et al.* Interaction-Driven Spectrum Reconstruction in Bilayer Graphene. *Science* **333**, 860-863 (2011).
39. Mayorov, A. S. *et al.* How Close Can One Approach the Dirac Point in Graphene Experimentally? *Nano Lett.* **12**, 4629-4634 (2012).
40. Fradkin, E. Critical behavior of disordered degenerate semiconductors. II. Spectrum and transport properties in mean-field theory. *Phys. Rev. B* **33**, 3263-3268 (1986).
41. Katsnelson, M. I. Zitterbewegung, chirality, and minimal conductivity in graphene. *Eur. Phys. J. B* **51**, 157-160 (2006).
42. Tworzydło, J., Trauzettel, B., Titov, M., Rycerz, A. & Beenakker, C. W. J. Sub-Poissonian Shot Noise in Graphene. *Phys. Rev. Lett.* **96**, 246802 (2006).
43. Das Sarma, S., Adam, S., Hwang, E. H. & Rossi, E. Electronic transport in two-dimensional graphene. *Rev. Mod. Phys.* **83**, 407-470 (2011).
44. Lucas, A., Crossno, J., Fong, K. C., Kim, P. & Sachdev, S. Transport in inhomogeneous quantum critical fluids and in the Dirac fluid in graphene. *Phys. Rev. B* **93**, 075426 (2016).
45. Bandurin, D. A. *et al.* Negative local resistance caused by viscous electron backflow in graphene. *Science* **351**, 1055-1058 (2016).
46. Levitov, L. & Falkovich, G. Electron viscosity, current vortices and negative nonlocal resistance in graphene. *Nat. Phys.* **12**, 672-676 (2016).
47. Tombros, N. *et al.* Large yield production of high mobility freely suspended graphene electronic devices on a polydimethylglutarimide based organic polymer. *J. Appl. Phys.* **109**, 093702 (2011).
48. Ki, D.-K., Fal'ko, V. I., Abanin, D. A. & Morpurgo, A. F. Observation of Even Denominator Fractional Quantum Hall Effect in Suspended Bilayer Graphene. *Nano Lett.* **14**, 2135-2139 (2014).
49. Grushina, A. L. *et al.* Insulating state in tetralayers reveals an even–odd interaction effect in multilayer graphene. *Nat. Commun.* **6**, 6419 (2015).

50. Nam, Y., Ki, D.-K., Koshino, M., McCann, E. & Morpurgo, A. F. Interaction-induced insulating state in thick multilayer graphene. *2D Mater.* **3**, 045014 (2016).